\documentclass[preprint,1p,11pt]{style/ISAS_IR} 

\usepackage[utf8]{inputenc}
\usepackage[T1]{fontenc}
\usepackage[per-mode=symbol]{siunitx} 
\usepackage{subfig}
\usepackage{tensor} % Hoch/Tief Indizes auch links von Variable
\usepackage{nicefrac} % for fractions with slash: a/b
\usepackage{comment} % block comment: \begin{comment} ... \end{comment}

\usepackage{microtype} % Optischer Randausgleich
\usepackage{lmodern} % scalable font 

\usepackage{amsmath,amssymb,amsfonts}
\usepackage[ruled]{algorithm2e} % ,vlined
\usepackage{etoolbox}
\makeatletter
\patchcmd{\@algocf@start}% <cmd>
{-1.5em}% <search>
{0pt}% <replace>
{}{}% <success><failure>
\makeatother

\usepackage{graphicx}
\usepackage{url}

\usepackage{mathtools}
\mathtoolsset{showonlyrefs}

\allowdisplaybreaks

\usepackage{acro}

\DeclareAcronym{TOA}		{short={TOA},		    long={time of arrival}, long-plural-form={times of arrival}}
\DeclareAcronym{TDOA}		{short={TDOA},		    long={time difference of arrival}, long-plural-form={time differences of arrival}}
\DeclareAcronym{DOTA}		{short={DOTA},		    long={difference of times of arrival}, long-plural-form={differences of times of arrival}, short-plural-form={DOTAs}}
\DeclareAcronym{ROTA}		{short={ROTA},		    long={round trip time of arrival}, long-plural-form={round trip times of arrival}}
\DeclareAcronym{LTDOA}		{short={L\hbox{-}TDOA},	long={local time difference of arrival}, long-plural-form={local time differences of arrival}}
\DeclareAcronym{LDOTA}		{short={L\hbox{-}DOTA},	long={local differences of times of arrival}, long-plural-form={local differences of times of arrival}}
\DeclareAcronym{DOTD}		{short={DOTD},		    long={difference of time differences}}
\DeclareAcronym{TTT}		{short={TTT},		    long={target transmission time}}
\DeclareAcronym{TDOT}		{short={TDOT},		    long={time difference of transmissions}, long-plural-form={time difference of transmissions}}
\DeclareAcronym{SSR}		{short={SSR},	    	long={secondary surveillance radar}}
\DeclareAcronym{PSR}		{short={PSR},	    	long={primary surveillance radar}}
\DeclareAcronym{LORAN}		{short={LORAN},	    	long={long range navigation system}}
\DeclareAcronym{MLAT}		{short={MLAT},	    	long={multilateration}}
\DeclareAcronym{MLE}		{short={MLE},	    	long={maximum likelihood estimator}}
\DeclareAcronym{RSS}		{short={RSS},	    	long={received signal strength}}
\DeclareAcronym{AOA}		{short={AOA},	    	long={angle of arrival}}
\DeclareAcronym{ML}		    {short={ML},	    	long={maximum likelihood}}
\DeclareAcronym{MAP}	    {short={MAP},	    	long={maximum a posteriori}}
\DeclareAcronym{MMSE}		{short={MMSE},	    	long={minimum mean square error}}
\DeclareAcronym{RMSE}		{short={RMSE},	    	long={root mean square error}}
\DeclareAcronym{LS}		    {short={LS},	    	long={least squares}}
\DeclareAcronym{LM}		    {short={LM},	    	long={Levenberg-Marquardt}}
\DeclareAcronym{GNSS}		{short={GNSS},	    	long={global navigation satellite systems}}
\DeclareAcronym{GPS}		{short={GPS},	    	long={Global Positioning System}}
\DeclareAcronym{ADSB}		{short={ADS-B},	    	long={automatic dependent surveillance -- broadcast}}
\DeclareAcronym{TOF}		{short={TOF},	    	long={time of flight}}
\DeclareAcronym{LORAWAN}	{short={LoRaWAN},	  	long={Long Range Wide Area Network}}
\DeclareAcronym{3DUSCT}	    {short={3D-USCT},	  	long={3D-Ultrasound Computer Tomography}}
\DeclareAcronym{EKF}		{short={EKF},	    	long={Extended Kalman Filter}}
\DeclareAcronym{UKF}		{short={UKF},	    	long={Unscented Kalman Filter}}
\DeclareAcronym{SDR}		{short={SDR},	    	long={software--defined radio}}
\DeclareAcronym{CPU}		{short={CPU},	    	long={central processing unit}}
\DeclareAcronym{CRLB}		{short={CRLB},	    	long={Cramér--Rao lower bound}}
\DeclareAcronym{VR}		    {short={VR},	    	long={virtual reality}}
\DeclareAcronym{SME}		{short={SME},	    	long={Symmetric Measurement Equation}}
\DeclareAcronym{LCD}		{short={LCD},	    	long={Localized Cumulative Distribution}}
\DeclareAcronym{EM}	    	{short={EM},	    	long={expectation--maximization}}
\DeclareAcronym{S2KF}	    {short={S$^2$KF},	    long={Smart Sampling Kalman Filter}}
\DeclareAcronym{ODE}	    {short={ODE},	    	long={ordinary differential equation}}
\DeclareAcronym{GM}	        {short={GM},	    	long={Gaussian mixture}}
\DeclareAcronym{DM}	        {short={DM},	    	long={Dirac mixture}}
\DeclareAcronym{LRKF}	    {short={LRKF},	    	long={linear regression Kalman filter}}
\DeclareAcronym{CoDiCo}	    {short={CoDiCo},	    long={coupled discrete and continuous densities}} 
\DeclareAcronym{vMF}		{short={vMF},	    	long={von Mises--Fisher}}
\DeclareAcronym{DMD}		{short={DMD},	    	long={Dirac mixture density}}
\DeclareAcronym{PDF}		{short={PDF},	    	long={probability density function}}
\DeclareAcronym{CDF}		{short={CDF},	    	long={cumulative density function}}
\DeclareAcronym{PCD}		{short={PCD},	    	long={projected cumulative distribution}}

\def\vec#1{\underline{#1}}

\def\mat#1{{\mathbf #1}}

\def\1_2{{\frac{1}{2}}}

\def\dd{{\,\operatorfont{d}}} % "d" operator for integration
\def\T{ ^\top } % Transpose 
\def\op#1{{\operatorfont{#1}}}

\newcommand{\rvec}[1]{\underline{\pmb{#1}}} % random vector
\newcommand{\rv}[1]{{\pmb{#1}}} % random variable

\def\ds{\displaystyle}

\def\NewR{\mathbb{R}} % {{\rm I\hspace{-.17em}R}}
\def\Eq#1{(\ref{#1})}

\def\Sec#1{Sec.~\ref{#1}}

\def\Fig#1{Fig.~\ref{#1}}

\def\Alg#1{Alg.~\ref{#1}}

\usepackage{color}

\newcommand\paren[1]{\left( #1 \right)}              % \paren{a}     (a)   (normal parentheses)
\newcommand\braces[1]{\left\lbrace #1 \right\rbrace} % \braces{a}    {a}
\newcommand\abs[1]{\left| #1 \right|}                % \abs{a}       |a|
\newcommand\hint[1]{ \quad \left\vert \; #1 \right.} % \hint{x^2}    |hint (on the right of an equation)

\SetKw{Fun}{Function} 
\SetKwFunction{KwSamplePrior}{samplePrior} 
\SetKwFunction{KwRound}{round} 
\SetKwFunction{KwGetStdNormalSamples}{stdNormalSamples} 
\SetKwFunction{KwChol}{chol} 
\SetKwFunction{KwNormalizeLOne}{normalizeL1} 
\SetKwFunction{KwMin}{min} 

\SetKwFunction{KwSampleProjected}{sample1D} 
\SetKwFunction{KwSampleSTwo}{sampleS1} 
\SetKwFunction{KwCumtrapz}{cumtrapz} 
\SetKwFunction{KwFindAdjacent}{adjacent} 
\SetKwFunction{KwRoots}{roots} 
\SetKwFunction{KwSelectBest}{select\_best} 
\SetKwFunction{KwSort}{sort} 
\SetKwFunction{KwProjectOrth}{project\_orth} 
\SetKwFunction{KwProject}{project} 
\SetKwFunction{KwRand}{rand} 
\SetKwFunction{KwAngToSTwo}{ang2S2} 
\SetKwFunction{KwAtanTwo}{atan2} 
\SetKwFunction{KwBackproject}{backproject}

\def\s { \hat{\vec x} }
\def\u { {\vec u} }
\def\x { {\vec x} } 

\def\fRef{ { \tensor*[^{\operatorfont{C\!}}]{f}{} } }
\def\fDM { { \tensor*[^{\operatorfont{DM\!}}]{f}{} } }
\def\FRef{ { \tensor*[^{\operatorfont{C\!}}]{F}{} } }

\begin{document}

\begin{frontmatter}
	
\title{ %\bf
Deterministic Sampling on the Circle \\
using Projected Cumulative Distributions
}

\author{Daniel~Frisch {\normalfont and} Uwe~D.~Hanebeck}
\address{Intelligent Sensor-Actuator-Systems Laboratory (ISAS)\\
Institute for Anthropomatics and Robotics\\
Karlsruhe Institute of Technology (KIT), Germany \\
e-mail: {\normalfont \texttt{daniel.frisch@ieee.org, uwe.hanebeck@ieee.org}}}

\begin{abstract}
%
% What we provide
%
We propose a method for deterministic sampling of arbitrary continuous angular density functions. 
%
% Why Deterministic?
%
With deterministic sampling, good estimation results can typically be achieved with much smaller numbers of samples compared to the commonly used random sampling. 
%
% UKF
%
While the Unscented Kalman Filter uses deterministic sampling as well, it only takes the absolute minimum number of samples. Our method can draw arbitrary numbers of deterministic samples and therefore improve the quality of state estimation. 
%
% Distance Measure
%
Conformity between the continuous density function (reference) and the Dirac mixture density, i.e., sample locations (approximation) is established by minimizing the difference of the cumulatives of many univariate projections. In other words, we compare cumulatives of probability densities in the Radon space. 
\end{abstract}
\end{frontmatter}
\acresetall

\section{Introduction}
\paragraph*{Context} 
%
% State Est & Control : Samples for nonlinear 
%
State estimation or control techniques for nonlinear systems often use samples (or particles) to represent the occurring densities. 

%
% Random sampling 
%
Obtaining discrete samples (on continuous domains) from continuous \acp{PDF} is therefore an important module in many state estimators and controllers. The ``brute force'' approach, often used to obtain ground truth for reference, is Monte Carlo Sampling with large numbers of random samples. There are universal but rather slow random sampling methods \cite{Hastings1997} and faster methods specialized for certain densities like the von Mises-Fisher distribution \cite{SDF15_Kurz}. 
 
%
% Deterministic sampling
%
In embedded systems subject to real-time constraints and limited memory, the number of samples should be rather small. With deterministic samples (instead of stochastic samples), comparable results can be achieved with much fewer samples. 

%
% Applications
%
Applications for deterministic sampling and filtering particularly in directional statistics include predictive control \cite{ACC15_Kurz}, heart phase estimation \cite{Fusion15_Kurz}, wavefront orientation estimation \cite{SDF18_Li}, and visual SLAM \cite{Fusion19_Bultmann}. 

\begin{figure}[t]  
	\centering 
	\includegraphics[width=0.6\linewidth] {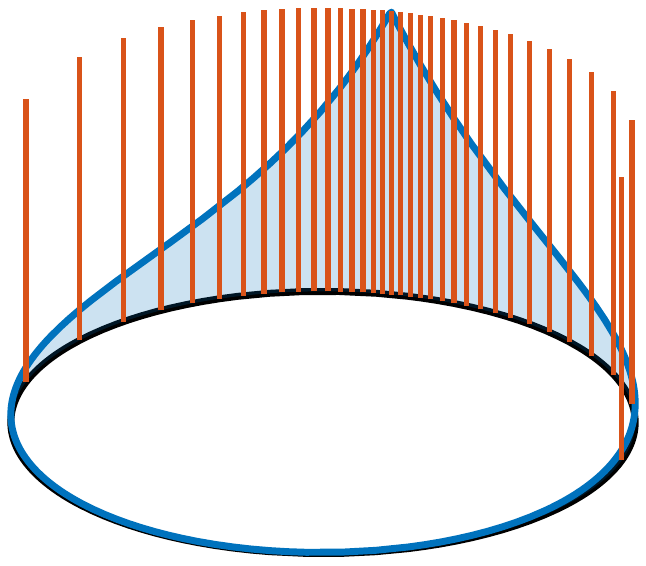} 
	\caption{ Wrapped Laplace Distribution (blue) on the circular domain (black), with proposed deterministic sampling result for 35 samples (red). }  
	\label{fig:eyecatcher} 
	% Generating Code: ISAS/Projects/RadonSampling/Matlab/WArbitraryS1.m  
\end{figure}

\paragraph*{Considered Problem} 
%
% Deterministic Sampling on Circle 
%
In this work we consider the problem of deterministic sampling of arbitrary continuous densities on the circular domain with an arbitrary number of samples. 

\paragraph*{State-of-the-art} 
%
% UKF
%
The minimalistic and popular deterministic sampling method of normal densities in the Euclidean domain is the basis of the \ac{UKF} \cite{julier1997new,julier02}. 
%
% Circle-UKF, 3 samples
%
The efficient concept of the \ac{UKF} has successfully been transferred to the circular domain \cite{ACC13_Kurz,ACC14_Kurz}, however inheriting equivalent limitations (only three samples, specific types of densities). 
%
% Circle-UKF, 5+ samples
%
Using higher order moments, the number of samples can be increased to five \cite{AES16_Kurz} and multiples of five with superposition techniques \cite{JAIF16_Kurz}. For specific densities, sampling based on the \ac{CDF} has been proposed \cite{JAIF16_Kurz} but is not invariant w.r.t. interval choice. 

%
% Equidistant
%
Weighted samples in an equidistant grid are very well suited for the circular domain \cite{MFI16_Kurz}, the sphere \cite{IFAC20_Pfaff}, and the torus \cite{MFI20_Pfaff} but expensive to extend to a high number of dimensions.
%
% Higher dimensions
%
\ac{UKF}-like sampling methods by contrast are applied to higher-dimensional directional estimation for orientations on the hypersphere \cite{TAC16_Gilitschenski,SPL16_Kurz}, for multivariate circular estimation on the torus \cite{TAES17_Kurz}, and for dual quaternions on special Euclidean groups \cite{IFAC20_Li-UPF,LCSS21_Li}, general Lie groups \cite{Brossard2017}, or arbitrary Riemannian manifolds \cite{Hauberg2013, Menegaz2019} -- all without exponential increase of computational cost.

%
% LCD - Euclidean
%
How can we make deterministic sampling more flexible, i.e., provide more samples than \ac{UKF}-like schemes, but avoid Cartesian products? One way to achieve this is based on the \ac{LCD} and a modified Cramér-von Mises distance. The \ac{LCD} transforms any density (either continuous or \ac{DM}) to a continuous representation via kernel convolution. The modified Cramér-von Mises distance is basically an $L^2$ norm of the difference of densities \cite{Izenman1991} but additionally averages over all kernel widths. \ac{LCD} and modified Cramér-von Mises distance together yield a distance measure between continuous and \ac{DM} densities in any combination \cite{MFI08_Hanebeck-LCD}, which has been successfully applied in the Euclidean domain \cite{AT15_Hanebeck}, especially for Gaussian densities \cite{JAIF14_Steinbring-S2KF,MFI20_Frisch}. 

%
% LCD - Directional
%
Early adaptions to directional estimation applied the \ac{LCD} in the Euclidean tangent space of the density's mean, placing samples on the coordinate axes only \cite{ECC19_Li} or distributing them in the entire tangent space \cite{Fusion19_Li}. Direct application of the \ac{LCD} on non-Euclidean manifolds has been performed for sample reduction (\ac{DM} to \ac{DM} comparison) on the sphere \cite{IFAC20_Frisch} and for dual quaternion sample reduction in the special Euclidean group SE(2) \cite{FUSION20_Li}. 
%
% LCD - Problems
%
Unfortunately, this method cannot easily be applied to arbitrary density functions and manifolds, because the involved integrals often do not exist in closed form. 

%
% PCD? 
%

%
% Star-like sampling
%
For the special case of the von Mises-Fisher density there is also a very efficient deterministic sampling method that places samples on an arbitrary number of ``beams'' in a star-like arrangement \cite{MFI20_Li}. It is very fast and more flexible than UKF-like, but the star-like arrangement doesn't always cover the state space homogeneously and purely according to the density function. 

\paragraph*{Contribution} 
%
% Arbitrary densities on Circle
%
In this paper, we present a method to optimally approximate a continuous angular density function $\fRef(\x)$ on the circular domain with a \ac{DMD} $\fDM(\x)$ with an arbitrary number of samples.

%%%%%%%%%%%%%%%%%%%%%%%%%%%%%%%%%%%%%%%%%%%%%%%%%%%%%%%%%%%%%%%%%%%%%%%%%%%%%%%%%%%%%%%%%%%%%%%%%%%%

\section{Overview}
\paragraph*{Key Idea}
%
% PCD Sampling on Circle 
%
We propose to extend the \ac{PCD} from the Euclidean space $\NewR^d$ \cite{CISS20_Hanebeck} to the circular domain $S^1$ and use it for deterministic sampling. 
%
% Projections: Univariate
%
By projecting to one-dimensional marginal distributions, we reduce multivariate problems to a set of univariate ones. 
%
% Unique CDF
%
In the univariate setting, cumulative distributions are uniquely defined and can easily be approximated even for arbitrary density functions. 

%
% Multiple Projections
%
In other words, we match a continuous density with a \ac{DMD} in the Radon domain. To optimally capture and transfer all of the density's details, it is important to include many different projections, which we implement in an iterative manner. 

\paragraph*{Problem Formulation} 
%
% Reference
%
$\fRef(\x),\; \x\in S^1$ is an arbitrary continuous density function on the circle, considered as reference density here. 
%
% DMA 
%
The goal is to obtain a \ac{DMD} 
\begin{align}
	\fDM(\x) &= \frac{1}{L} \sum_{i=1}^L \delta(\x-\s_i) 
\end{align}
with sample locations $\s_i \in S^1\,, \;\, i\in\braces{1,2,\hdots,L}$. 
%
% Equality 
%
This \ac{DMD} should optimally approximate the given continuous reference density, limited in accuracy only by the allowed number of samples $L$. 
%
% Inputs & Outputs
%
Required inputs are 
\begin{itemize} 
	\item[I1] the number $L$ of wanted samples,
	\item[I2] a numerical function handle of a continuous angular reference density function  $\fRef(\x),\; \x\in S^1$. 
\end{itemize} 
Obtained outputs are the sample locations $\s_i \in S^1$. 

\section{Projection of the Circular Domain} 
\label{sec:proj}
%
% Projection Formulation 
%
Projection along a certain direction $\u \in S^1$ allows to compare one-dimensional \acp{PDF} $f(r|\u)$ at a time. \Acp{CDF} $F(r|\u)$ are uniquely defined in one dimension and can also be easily calculated from the \acp{PDF} via the trapezoidal rule with proposal samples (if no closed-form solution is available). Furthermore it is easy to compare two one-dimensional \acp{CDF}. 

%
% 2 Considered Projections
%
The following two types of projections $f(r|\u)$ of circular densities $f(\x)$, exponential map and orthographic projection, appear to be equally convenient for our purpose. 

\subsection{Exponential Map} 
%
% Cut Circle Open
%
Consider the circular domain as a real interval of length $2 \pi$ by cutting the unit circle open at an arbitrary position $\u\in S^1$  
\begin{align}
	f(r|\u) &= \begin{cases}
	    f\!\paren{ \begin{bmatrix} \cos(r-\angle\u) \\ \sin(r-\angle\u) \end{bmatrix} }\;, & 0 \leq r \leq 2\pi \;, \\
	    0\;, & \text{otherwise}  \;, \end{cases} \label{eq:proj:exp}
\end{align}
where $\angle\u = \op{atan2}(\u^{(2)},\, \u^{(1)})$ is an angular representation of $\u$.

\begin{figure}[t]  
	\centering 
	\includegraphics[width=0.6\linewidth] {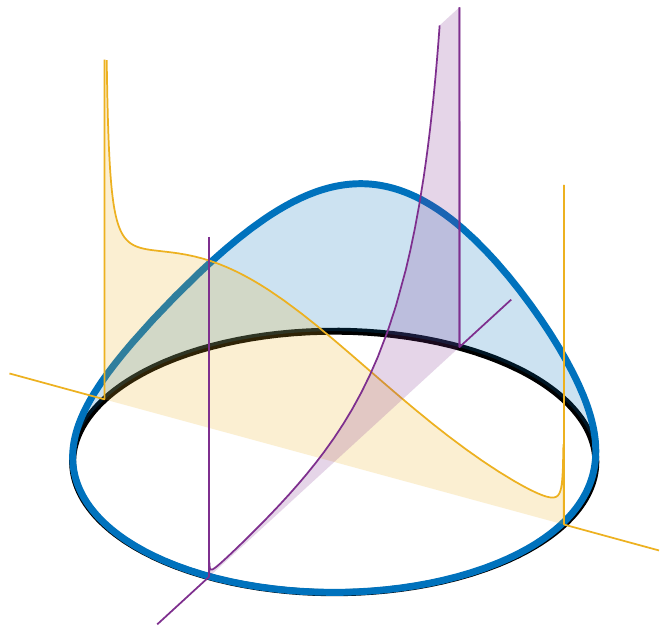} 
	\caption{ Continuous density function (blue) and two orthographic projections or marginals (yellow, purple), see \Eq{eq:proj:orth}.  }  
	\label{fig:OrthProj} 
	% Generating Code: ISAS/Projects/RadonSampling/Matlab/WArbitraryS1.m  
\end{figure}

\subsection{Orthographic Projection} 
%
% Projection in R² 
%
Consider the Euclidean embedding of the circular manifold $S^1$ in  $\NewR^2$. We then perform a linear projection using the direction vector $\u$ 
\begin{align}
	\rv r   &= \u \T \rvec x \enspace, 
\end{align}
yielding a univariate random variable $\rv r$. 
In terms of densities, we calculate the marginal distribution along $\u$
\begin{align}
	f(r|\u) &= \int_{S^1} f(\x) \, \delta(r - \u\T \x) \dd \x \hint \alpha=\angle\x \\
	%&= \int_{\alpha=0}^{2\pi} f\!\paren{ \begin{bmatrix} \cos(\alpha) \\ \sin(\alpha) \end{bmatrix} }  \delta\!\paren{ r - \u \T \begin{bmatrix} \cos(\alpha) \\ \sin(\alpha) \end{bmatrix} } \dd \alpha \\ 
	&= \int\limits_{\alpha=0}^{2\pi} 
	\!\!f\!\paren{ \begin{bmatrix} \cos(\alpha) \\ \sin(\alpha) \end{bmatrix} }  
	\delta\!\paren{\! r -\! \begin{bmatrix} \cos(\angle \u) \\ \sin(\angle \u) \end{bmatrix}^{\!\!\top}\!\!\! \begin{bmatrix} \cos(\alpha) \\ \sin(\alpha) \end{bmatrix} } \!\operatorfont{d} \alpha \\ 
	&= \int\limits_{\alpha=0}^{2\pi} 
	f\!\paren{ \begin{bmatrix} \cos(\alpha) \\ \sin(\alpha) \end{bmatrix} }  
	\delta\!\paren{ r - \cos(\alpha - \angle \u)} \dd \alpha \\ 
	&= \begin{cases} 
	\ds \sum_{i=1}^2 f\!\paren{ \begin{bmatrix} \cos(\alpha_i+\angle \u) \\ \sin(\alpha_i+\angle \u) \end{bmatrix} }\! \frac{1}{\abs{\sin(\alpha_i)}}\,, & \abs{r}\leq1\,, \\
	%\operatorfont{undef}\,, & \abs{r}=1 \,, \\
	0\,, & \abs{r} > 1 \,,\end{cases} \label{eq:proj:orth} 
\end{align}
with
\begin{align} 
	\alpha_i &= \begin{cases} \arccos(r)\,, & i=1 , \\ 2\pi-\arccos(r)\;, & i=2 \enspace. \end{cases} 
\end{align}
See \Fig{fig:OrthProj}  for a visualization of two orthographic projections.

\begin{figure}[t]  
	\centering 
	\includegraphics[width=0.7\linewidth] {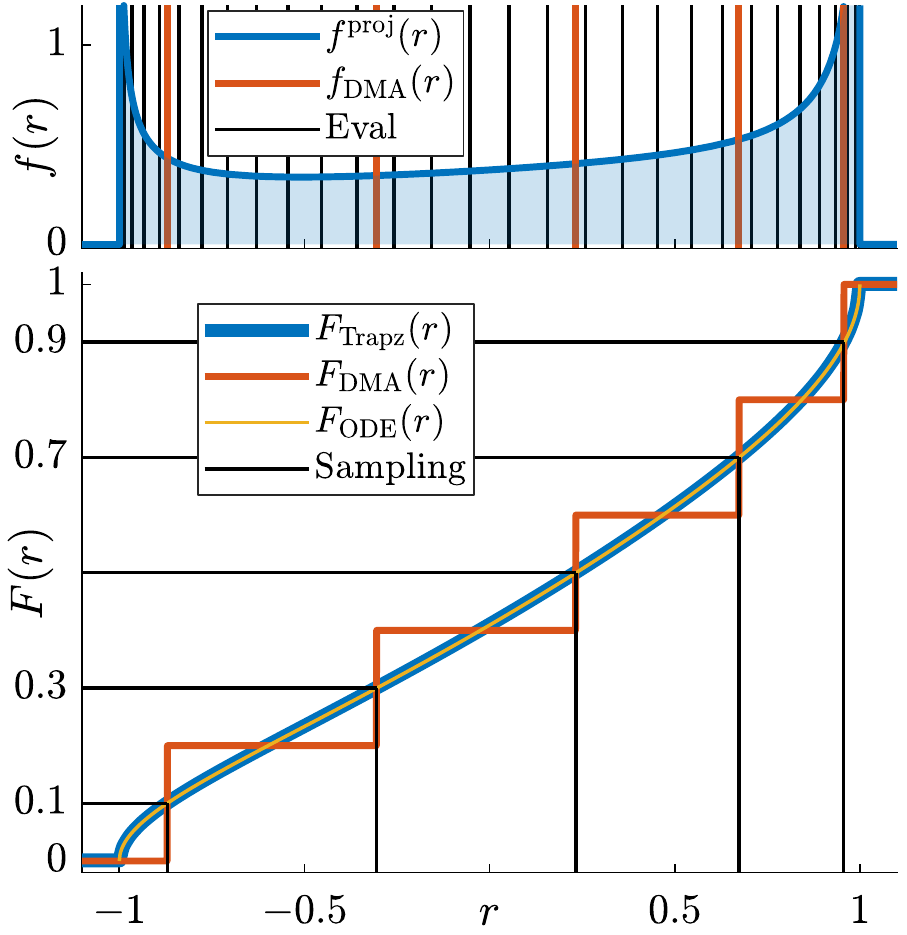} 
	\caption{ Procedure for deterministic sampling of a projected von Mises-Fisher density, using orthographic projection \Eq{eq:proj:orth}. Upper part: we evaluate $f(r)$ (blue) at the fixed evaluation points $t_j^{\op{h}}$ (black) as well as previous sample locations (red). Lower part: Trapezoidal integration on said evaluation points is performed (blue). Compare the ground truth obtained with a numerical ODE solver (yellow). Then, one-dimensional deterministic sampling is performed (black), yielding an approximating \ac{DM} distribution function (red). See also \Alg{alg:sampleProjected} for a more detailed description.  }  
	\label{fig:DetSampling} 
	% Generating Code: ISAS/Projects/RadonSampling/Matlab/WArbitraryS1Proj1.m
\end{figure}

\section{Implementation} 
\label{sec:Implementation}
With a suitable projection at hand, we can now start approximating the continuous density. 
%
% CDF 
%
It is well known that samples of any one-dimensional density, like our projected \ac{PDF}, can easily be drawn when the inverse of the \ac{CDF} is available. 
%
% Overview
%
Therefore, we seek to obtain the following intermediate results one by one in the course of this section: 
\begin{itemize}
    \item reference \ac{PDF} $\fRef(\x)$ (is given), 
    \item projected \ac{PDF} $\fRef(r|\u)$,
    \item projected \ac{CDF} $\FRef(r|\u)$,
    \item inverse \ac{CDF} $\FRef^{-1}(p|\u)$, 
    \item sample locations $r_i$, 
    \item sample updates $\Delta \x_i$.  
\end{itemize}
%
% Iterative update
%
The procedure will then be repeated iteratively for different projections $\u$.

\begin{algorithm}[t] 
	\DontPrintSemicolon 
	\caption{ Calculate sample steps that make a \ac{DMD} approximate a continuous density by matching the cumulatives, in the univariate (projected) setting. }  
	\label{alg:sampleProjected}  
	\Fun{} \\ 
	$\braces{ \Delta r_i }_{i=1}^{L} \gets$ \KwSampleProjected{ $ f_{\rv r}(\cdot),\; \braces{ r_i }_{i=1}^{L} \;$ } \\ 
	\KwIn{  
		$f_{\rv r}(\cdot)$: continuous reference density in one dimension, \newline  
		$\braces{ r_i }_{i=1}^{L}$: current sample approximation  
	} 
	\KwOut{  
		$\braces{ \Delta r_i }_{i=1}^{L}$: proposed step for each sample, to improve similarity to $f_{\rv r}(\cdot)$   
	} 
	$\braces{t_i^{\op{h}}}_{i=1}^{L^{\op{h}}}$ \tcp*{Fixed evaluation points}  
	$\braces{t_j}_{j=1}^{L^{\op{e}}=L^{\op{h}}+L}  
	\gets \braces{t_i^{\op{h}}}_{i=1}^{L^{\op{h}}}    \;\cup\;    \braces{ r_i }_{i=1}^{L}$ \; 
	$\braces{F_j}_{j=1}^{L^{\op{e}}} 
	\gets $ \KwCumtrapz{ $ \braces{t_j}_{j=1}^{L^{\op{e}}},\; \braces{f_{\rv r}(t_j)}_{j=1}^{L^{\op{e}}} \;$ } \;  
	$\braces{F_j}_{j=1}^{L^{\op{e}}} \gets  \braces{F_j + \frac{1-F_{L^{\op{e}}}}{2} }_{j=1}^{L^{\op{e}}}$ \tcp*{Centering} 
	\For{ $i \gets 1 $ \KwTo $L$ } { 
		$F^{\op{det}} \gets \frac{2 i-1}{2L} $ \tcp*{Deterministic sampling} 
		$\paren{ j^{\op{L}}, j^{\op{R}} } \gets $ \KwFindAdjacent{ $F^{\op{det}} ,\; \braces{F_j}_{j=1}^{L^{\op{e}}}\; $} \;
		\tcp{Quadratic interpolation} 
		$m \gets \frac{f_{j^{\op{R}}}-f_{j^{\op{L}}}}{t_{j^{\op{R}}}-t_{j^{\op{L}}}}$ \;
		$(a,b,c) \gets F_{j^{\op{L}}} + \int_{t_{j^{\op{L}}}}^x m \cdot \paren{x-t_{j^{\op{L}}}} \dd x  \overset{!}{=} F^{\op{det}}$   \;
		$(x_1^{\op{quad}},x_2^{\op{quad}}) \gets$ \KwRoots{ $a, b, c \; $ } \;
		\tcp{Linear interpolation} 
		$x^{\op{lin}} \gets \frac{F^{\op{det}}-F_{j^{\op{L}}}}{m} + t_{j^{\op{L}}}$ \; 
		\tcp{Updated sample location} 
		$r_i^{\op{e}} \gets $ \KwSelectBest{ $ x_1^{\op{quad}} ,\; x_2^{\op{quad}} ,\; x^{\op{lin}} \; $ } \; 
	} 
	\tcp{Assign $r_i$ and $r_i^{\op{e}}$} 
	$\paren{ \braces{ r_i^{\op{sort}} }_{i=1}^{L} ,\; \braces{ j_i }_{i=1}^{L} } 
	\gets $ \KwSort{ $\braces{ r_i }_{i=1}^{L}\;$ } \;
	\For{ $i \gets 1 $ \KwTo $L$ } {
		$\Delta r_{j_i} \gets r_i^{\op{e}} - r_i^{\op{sort}} $ \tcp*{Sample step} 
	}
	% Code: /ISAS/Projects/RadonSampling/Matlab/lib/wassersteinSN.m
\end{algorithm}

\subsection{Composite Trapezoidal Integration} 
%
% Numerical Integration
%
The projected \ac{PDF} $\fRef(r|\u)$ is available in closed form by inserting the given $\fRef(\x)$ into \eqref{eq:proj:exp} or \eqref{eq:proj:orth}. Since we are permitting arbitrary density functions, a closed-form representation of the according \ac{CDF}  
\begin{align}
    \FRef(r|\u) &= \int_{t=-\infty}^r f(t|\u) \dd t 
\end{align}
is not possible in general. However, we know that the integrand $f(r|\u)$ has limited support, i.e., $r\in[0,2\pi]$ for the exponential map projection \eqref{eq:proj:exp}, and $r\in[-1,1]$ for the orthographic projection \eqref{eq:proj:orth}. 
%
% Trapezoidal
%
To obtain an approximation of $F(r|\u)$, we apply the composite trapezoidal rule with an adaptive set of function evaluation points $t_j$. 

%
% Fixed evaluation points
%
A fixed set of homogeneous function evaluation points $t_j^{\operatorfont{h}}$ inside the support interval is always used to ensure a good general approximation of the \ac{CDF}'s global shape. 
%
% Proposal evaluation points
%
Additionally, in order to maintain proper accuracy of the numerical integral even in the case of very localized \acp{PDF} with small extent, the projected samples $r_i^{\operatorfont{p}}$ in the currently assumed approximating density $\fDM(r|\u)$ are always included into the set of function evaluation points. 

%
% Piecewise quadratic CDF
%
Summarizing, after composite trapezoidal integration of $\fRef(r|\u)$ with said evaluation points, we now have a piecewise linear representation of the projected reference \ac{CDF}, $\FRef(r|\u)$.

\begin{algorithm}[!ht] 
	\DontPrintSemicolon 
	\caption{ \ac{PCD}-based deterministic sampling of conditional circular densities. }  
	\label{alg:detsamp}  
	\Fun{} $\braces{ \s_i }_{i=1}^{L} \gets$ \KwSampleSTwo{ $ f_{\rvec x}(\cdot) ,\; L \; $} \\ 
	\KwIn{  
		$f_{\rvec x}(\cdot)$: continuous circular density, $\x\in S^2$, \newline  
		$L$: number of wanted samples 
	} 
	\KwOut{  
		$\braces{ \s_i }_{i=1}^{L}$: deterministic samples on the circle that approximate $f_{\rvec x}(\cdot)$  
	} 
	$N \gets 2$ \tcp*{Projections per iteration} 
	\tcp{High quality for visualization} 
	$M \gets 200$ \tcp*{Number of iterations}
	$\lambda_0 \gets 0.99$ \tcp*{Update step decrease factor} 
	\tcp{Initialization}  
	$\lambda \gets 1$ \;
	$\braces{ \s_i }_{i=1}^{L} \gets $ \KwRand{ $L ,\; S^2 \, $ } \;
	%\tcp{Iterate}
	\For{ $m \gets 1 $ \KwTo $M$ }  { 
		$\varphi_0 \gets $ \KwRand{ $1 ,\; S^2 \, $ } \;
		$\braces{ \Delta \s_i}_{i=1}^{L}  \gets \vec 0 $ \; 
		\For{ $n \gets 1 $ \KwTo $N$ } {  
			\tcp{Symmetric projections} 
			$\varphi \gets \pi \cdot (n-1) / N + \varphi_0 $ \;
			$\u \gets \begin{bmatrix} \cos(\varphi) \\ \sin(\varphi) \end{bmatrix}$  \;
			\tcp{Project the samples $\s_i \rightarrow  r_i$} 
			$\braces{  r_i }_{i=1}^{L} \gets   \braces{ \u\T\s_i }_{i=1}^{L} $ \;
			\tcp{Project the density $f_{\rvec x}(\cdot) \rightarrow f_{\rv r}(\cdot|\u)$} 
			\tcp{according to \Sec{sec:proj}}
			$f_{\rv r}(\cdot) \gets $ \KwProject{ $ f_{\rvec x}(\cdot) ,\; \u \;$ } \; 
			\tcp{Get projected sample updates  }
			\tcp{using \Alg{alg:sampleProjected} } 
			$\braces{ \Delta r_i }_{i=1}^{L} \gets$ \KwSampleProjected{ $ f_{\rv r}(\cdot),\; \braces{ r_i }_{i=1}^{L} \; $ } \;
			\tcp{Get sample updates in $\NewR^2$} 
			$\braces{ \Delta \x_i}_{i=1}^{L}  \gets \{$ \KwBackproject{ $\Delta r_i\; $ } $\}_{i=1}^{L} $ \;
			$\braces{ \Delta \s_i}_{i=1}^{L}  \gets \braces{ \Delta \s_i + \Delta \x_i }_{i=1}^{L} $ \;
		} 
		$\lambda \gets \lambda \cdot \lambda_0$ \; 
		\For{ $i \gets 1 $ \KwTo $L$ } {
			\tcp{Perform sample update} 
			$\s_i \gets \s_i + \lambda \, \Delta \s_i / N  $ \; 
			\tcp{Restrict to $S^2$ }
			$\varphi_i \gets $ \KwAtanTwo{ $\s_i^{(2)} ,\; \s_i^{(1)} \;$  } \; 
			$\s_i \gets \begin{bmatrix} \cos(\varphi_i) \\ \sin(\varphi_i) \end{bmatrix} $	
		}
	} 
	% Code: /ISAS/Projects/RadonSampling/Matlab/WArbitraryS1.m 
\end{algorithm}

\subsection{Deterministic Sampling}
\label{sec:detsamp}
%
% Uniform Deterministic Samples 
%
We draw deterministic samples $p_i$ that are uniformly distributed in $[0,1]\,$, 
\begin{align}
    p_i &= \frac{2 i-1}{2L}\;, &  i&\in\braces{1,2,\hdots,L}\;,
\end{align}
% 
% Projected Deterministic Samples
%
and propagate them through the inverse \ac{CDF} to obtain deterministic samples $r_i$ of $\fRef(r|\u)$
\begin{align}
    r_i &= \FRef^{-1}(p_i|\u)\;, &  i&\in\braces{1,2,\hdots,L}\;. 
\end{align}

%
% Inverse CDF
%
Under the assumptions that have been made with the trapezoidal rule, our representation of $\fRef(r|\u)$ is piecewise linear, and thus $\FRef(r|\u)$ is a piecewise quadratic. 
Therefore, evaluation of $\FRef^{-1}(p_i|\u)$ for any $p_i$ to obtain $r_i$ involves two things. 
First, a search for the relevant interval, i.e., an adjacent pair $(t_{\op{L}}, t_{\op{R}})$ from the trapezoidal function evaluation points $t_i$ such that $\FRef(t_{\op{L}}|\u) \leq p_i < \FRef(t_{\op{R}}|\u)$. 
Second, the quadratic (or sometimes linear) function that represents the \ac{CDF} in this segment has to be inverted, what is easily done in closed form. 

Of course, if a closed-form representation of the projected \ac{CDF} or its inverse is available, we can use that directly for sampling, with no need for trapezoidal integration. For example, a fast approximation of the von Mises-Fisher density's cumulative (in conjunction with the exponential map) is available in closed form \cite{Hill1977}. 

%
% Projected Samples Available
%
At this point we have the deterministic sample locations $r_i$ in the projected space that is defined by the projection direction $\u$. 

%
% Figure
%
Compare \Fig{fig:DetSampling} for a visualization of \ac{CDF}-based sampling in the projected space.

\begin{figure*}[t] 
	\begin{center} 
		\subfloat[von Mises  \label{fig:eval:VM}] 
		{\includegraphics[width=.24\linewidth] {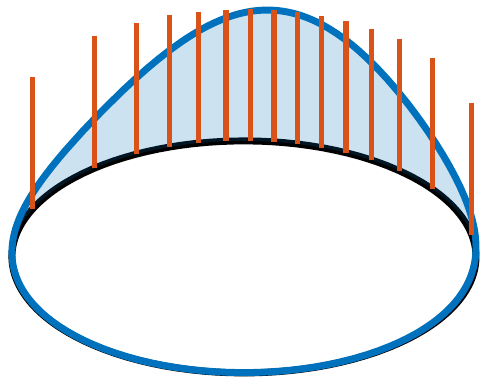}} \hfill 
		\subfloat[wrapped Cauchy  \label{fig:eval:WC}] 
		{\includegraphics[width=.24\linewidth] {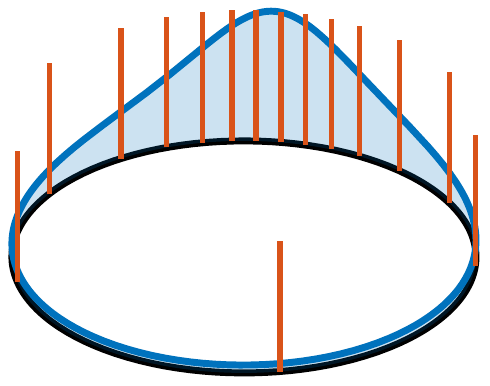}} \hfill 
		\subfloat[wrapped normal  \label{fig:eval:WN}] 
		{\includegraphics[width=.24\linewidth] {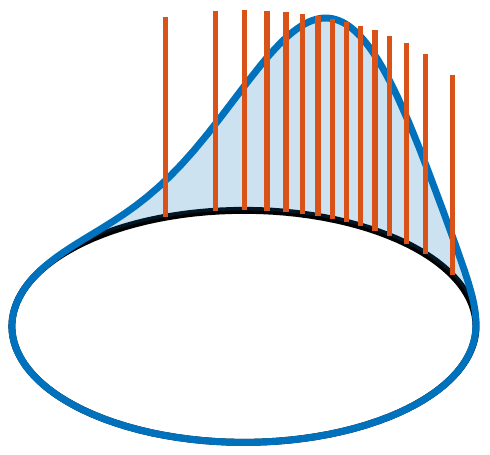}} \hfill 
		\subfloat[wrapped exponential  \label{fig:eval:WE}] 
		{\includegraphics[width=.24\linewidth] {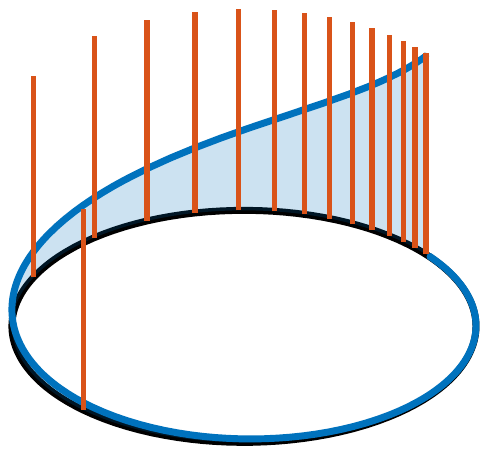}} \hfill 
		\subfloat[von Mises mixture  \label{fig:eval:Mix}] 
		{\includegraphics[width=.24\linewidth] {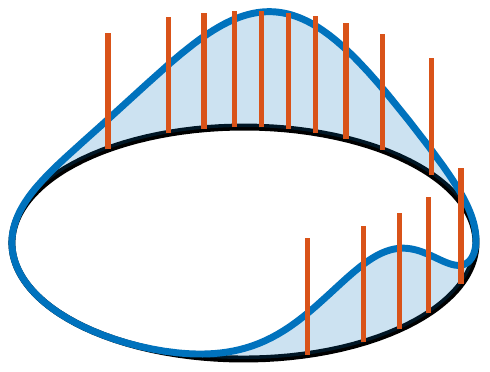}} \hfill 
		\subfloat[custom distribution, sinusoidal \label{fig:eval:Custom2}] 
		{\includegraphics[width=.24\linewidth] {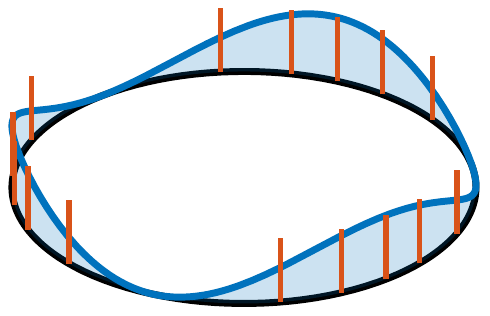}} \hfill 		
		\subfloat[piecewise constant \label{fig:eval:PWC}] 
		{\includegraphics[width=.24\linewidth] {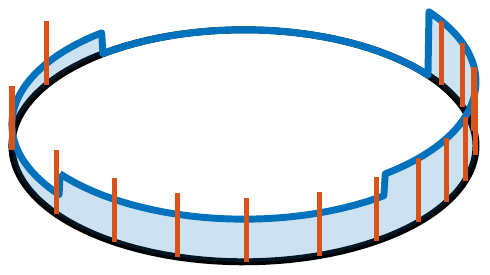}} \hfill 
		\subfloat[uniform \label{fig:eval:Uniform}] 
		{\includegraphics[width=.24\linewidth] {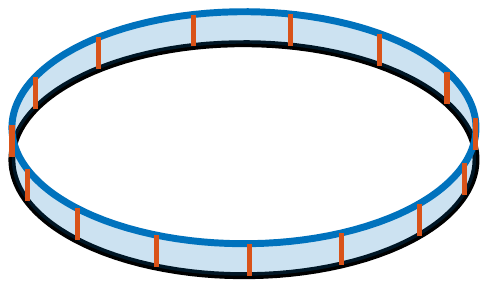}} \hfill 
	\end{center} 
	\caption{Illustration of various circular distributions and deterministic samples obtained with the proposed method. 
		Continuous probability density function (blue) on the angular domain $S^1$ (black), with sampling results (red). 
		For better visualization, the length of the red lines representing the unweighted samples has been set to the maximum density function value (mode) instead of the sample weight $1/L$.  \label{fig:eval} } 
	% Generating Code: /media/frisch/DATEN_SSD/ISAS/Projects/RadonSampling/Matlab/WArbitraryS1.m 
\end{figure*}

\subsection{Sample Update} 
% 
% Backpropagation
%
The projected sample locations $r_i$ now have to be backprojected to the original domain $S^1$. 
%
% Symmetric projections
%
We typically use updates from several symmetrically arranged projections simultaneously. 

%
% Association
%
The projected samples generated as described in \Sec{sec:detsamp} are not naturally associated with the existing samples from previous iterations. Thus, we have to find an appropriate association first. Projection also helps us here: in the one-dimensional case, the association that minimizes the global distance of associated point pairs can simply be obtained by element-wise comparison of the sorted sets. The according global distance is also called Wasserstein distance. 

%
% Pseudocode
%
Refer to \Alg{alg:sampleProjected} for a pseudocode representation of the procedure described in \Sec{sec:Implementation} up to here.

\begin{figure}[t] 
	\begin{center} 
		\subfloat[Fixed \label{fig:a2}] {\includegraphics[width=.3\linewidth] {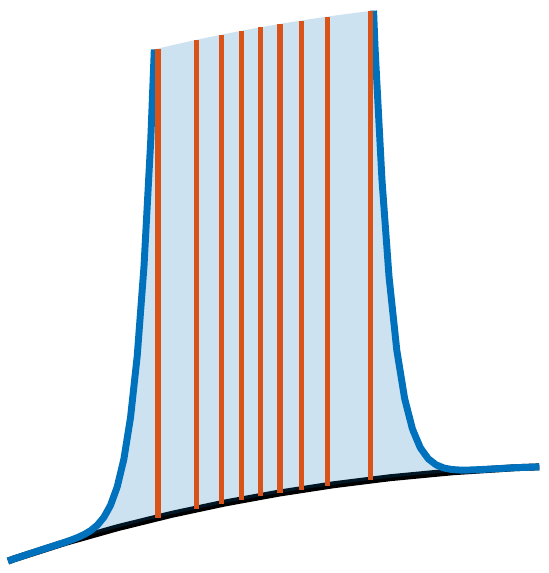}} 
		\subfloat[Adaptive \label{fig:a1}] {\includegraphics[width=.3\linewidth] {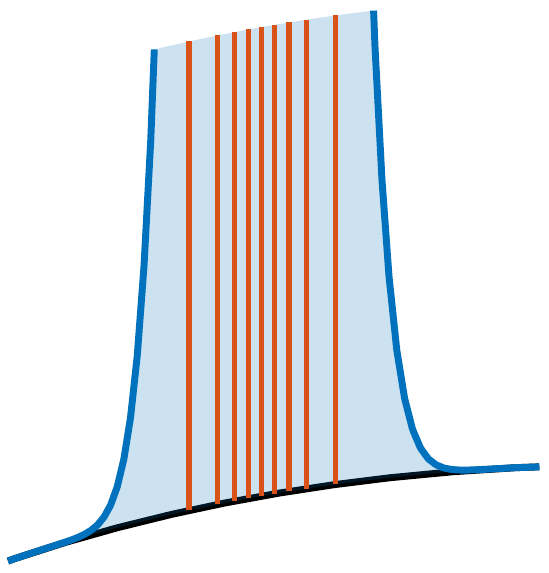}} 
		\hfill 
	\end{center} 
	\caption{ Deterministic circular sampling using (a) only a fixed set of 30 evaluation points $t_j^{\op{h}}$ versus (b) the 30 fixed points plus the previous samples, for better numerical integration. The difference for this quite ``narrow'' von Mises-Fisher distribution ($\kappa=500$) is notable. \label{fig:narrow}} 
	% Generating Code: ISAS/Projects/RadonSampling/Matlab/WArbitraryS1.m
\end{figure}

\subsection{Multiple Projections} 
%
% Symmetric
%
To equally consider all dimensions, we propose to use a symmetric set of $N$ projections in each iteration step. For $N=2$ projections per iteration, we choose projections that are orthogonal (\SI{90}{\degree} between them) but with random orientation, see \Fig{fig:OrthProj} for an example. 
%
% Averaged update
%
The individual sample updates from each projection are averaged, thus yielding the total update $\Delta \s_i$ of the current iteration step.

\subsection{Iterative Update} 
%
% Repeat
% 
The procedure is repeated until the arrangement of the samples obtains an acceptable quality. 
%
% Factor
%
In order to asymptotically reach a stationary state, we propose to multiply sample updates with an exponentially decreasing factor $\lambda$. This accounts for the fact that more and more information (from more projections) is already present in the sample locations, and the amount of extra information provided by every additional iteration decreases.  

%
% algorithm2e
%  
Refer to \Alg{alg:detsamp} for a more detailed presentation regarding the iterative sample update scheme.

\section{Evaluation} 
\label{sec:evaluation}
%
% Visual Examples
%
The flexibility of the proposed method is demonstrated by showing obtained deterministic samples form various different density functions, see \Fig{fig:eval}. 

%
% Adaptive Eval Points
% 
Our adaptive choice of evaluation points for numerical integration allows for an accurate approximation even for ``narrow'' densities, where fixed evaluation points alone would not be sufficient. See \Fig{fig:narrow} for an example.

\section{Conclusions} 
\label{sec:conclusion}
%
% Overview
%
We present a method to generate any number of deterministic samples for any continuous density function on the circle. 

%
% Computational Effort
%
It does not require gradient-based numerical optimization like \ac{LCD}-based methods. Instead, we use the trapezoidal rule with  adaptive support points on a given interval in an iterative method. 
%
% Minimalistic optimality measure
%
Furthermore, the distance measure is simple and undisputable: Matching the cumulatives is always an adequate solution for univariate densities. No parameters or weighting functions have to be chosen. With the help of the \ac{PCD}, we can apply the same elementary method (matching one-dimensional cumulatives) to higher dimensions. 

%
% Outlook
%
In the future, we will extend this method to higher-dimensional geometries such as the hypersphere and the torus. While calculating the projected density was easy on the circle, it will be more difficult in higher dimensions. We will look for closed-form solutions that work for specific types of densities. Furthermore, numerical integration techniques with an adaptive choice of evaluation points will be pursued and also pure sample reduction techniques, where no integration is necessary. 
%
% Projections
%
Presumably, orthographic projection is a good choice for hyperspherical higher-dimensional extensions of the circle, and the exponential map for the Cartesian product of circles, i.e., toroidal manifolds.

%\newpage
%
% Bibliography
%
%\vspace{.1cm}
\bibliographystyle{bib/IEEEtran}
\bibliography{bib/IEEEabrv, bib/ISASPublikationen, bib/paperlocal} % bib/ISASPublikationen_laufend, 

\end{document}